\documentclass{aastex62}         
\received{2019 August 27}
\revised{2019 October 31}
\accepted{2019 November 12}
\submitjournal{ApJ Letters}

\shorttitle{TESS observations of comet Wirtanen}
\shortauthors{Farnham et al.}

\begin{document}

\title{First Results from TESS Observations of Comet 46P/Wirtanen}

\correspondingauthor{Tony L. Farnham}
\email{farnham@astro.umd.edu}

\author[0000-0002-4767-9861]{Tony L. Farnham}
\affil{University of Maryland \\
Department of Astronomy \\
College Park, MD 20742,  USA}

\author[0000-0002-6702-7676]{Michael S. P. Kelley}
\affil{University of Maryland \\
Department of Astronomy \\
College Park, MD 20742,  USA}

\author[0000-0003-2781-6897]{Matthew M. Knight}
\affil{University of Maryland \\
Department of Astronomy \\
College Park, MD 20742,  USA}

\author[0000-0002-4230-6759]{Lori M. Feaga}
\affil{University of Maryland \\
Department of Astronomy \\
College Park, MD 20742,  USA}



\begin{abstract}

  We report on initial results from 20 days' worth of TESS spacecraft
  observations of comet 46P/Wirtanen.  The long-duration, high-cadence
  measurements show a 2018~September~26 outburst that exhibited a two-phase,
  0.5 mag brightening profile, and may be the best \added{temporally}
  characterized natural outburst ever recorded.  Gas velocities from the
  outburst peaked at 800~$\mathrm{m~s^{-1}}$, while dust expanded at only 10s
  of $\mathrm{m~s^{-1}}$.  Coadded images also revealed a previously
  unreported dust trail that extends beyond the 24\degr{} field of view.

\end{abstract}



\section{Introduction} \label{sec:intro}

The Transiting Exoplanet Survey Satellite (TESS) searches for extrasolar
planets by observing a sector of the sky for 27.4 days and using high-quality
photometric measurements of the stars in the field to look for transits
\citep{RickerEtal:TESS}.  A TESS sector covers 84\degr{} of ecliptic
\replaced{longitude}{latitude} (\replaced{$|l|>6$\degr{}}{$|\beta|>6$\degr{}}
) with four cameras, each with a 24\degr{} field of view (21-arcsec pixels).
Images are obtained every 2~minutes and then coadded into 30-minute exposures
that are saved as Full-Frame Images (FFI).  During its 27.4-day window, each
sector may produce as many as \replaced{1348}{1315} observations.  TESS will observe 26 of
these sectors (13 South and 13 North of the ecliptic plane) covering 90\% of
the sky during its 2-year primary mission.

Because of the large field-of-view, comets and asteroids serendipitously
appear in the TESS data.  Approximately 50~comets will be bright enough to be
detected in the two years of TESS observations, and we are using the
long-duration, high-cadence observations to monitor these comets for
rotational variability, outburst activity and coma studies.  The data are
ideal for exploring lightcurve behavior, because the 30-minute cadence
permits studies of short-term variability, while the 27-day sequences allow
measurements of periodicities $>$24~hrs (which are difficult to obtain from
ground-based observations) and provide the opportunity to search for changes
in the rotation rate \citep[e.g.,][]{BodewitsEtal:tgk}.  The TESS
observations are also well-suited for outburst studies, as the continuous
coverage and high-precision photometric capabilities allow searches for even
small events.  For any outbursts that occur during a 27-day sector window,
the 30-minute cadence will accurately document the start-time, brightening
profile, and peak brightness---characteristics that are rarely captured for
typical outburst detections---allowing for detailed comparisons that could
provide clues to the mechanisms at work in \replaced{these}{this} phenomenon.  Finally, the
wealth of observations allow for studies of the spatial structures in
cometary comae, including the ability to coadd large amounts of data (phased
to rotation periods if they are known) to investigate faint features that
might exist.

In its approach to perihelion, comet 46P/Wirtanen appeared within the TESS
field of view.  Wirtanen is a particularly interesting comet for
demonstrating TESS's potential contributions to cometary science. It is
a small hyperactive comet that made an historically close approach to the
Earth (0.077 AU) in December 2018, and is a potential target for future
spacecraft missions. We report here on first results from the observations of
a comet with TESS. 

\section{TESS Observations} \label{sec:observations}

During the 2018 TESS Sector~3 observations, Wirtanen appeared in the field of
view of Camera~2, CCD~\#3. From September~20 to October~17, 1288 FFI images
of the comet were acquired.  During this time, the comet's heliocentric
distance dropped from 1.51 to 1.30~AU and the geocentric distance from 0.58
to 0.36~AU, while the solar phase angle changed from 23.2\degr{} to
29.8\degr{}.  During the first and last three days of the
sector observations, the spacecraft was testing new pointing software,
increasing the jitter \citep{FausnaughEtal:TESS_sec3} and causing $\sim$340
observations to be considered ``non-science-quality''.  Other suboptimal
images include a sequence in which the scattered light from Earth was not
well removed and occasional frames that are smeared due to TESS momentum
dumps that occur every $\sim$2.5 days.  Although some of these data may be
recoverable, we have removed them for the initial analyses reported here.  In
total, we used 935 images spanning 20.4~days.

\section{Data Reduction and Analyses} \label{sec:reduction}

\begin{figure}[hbt]
\plotone{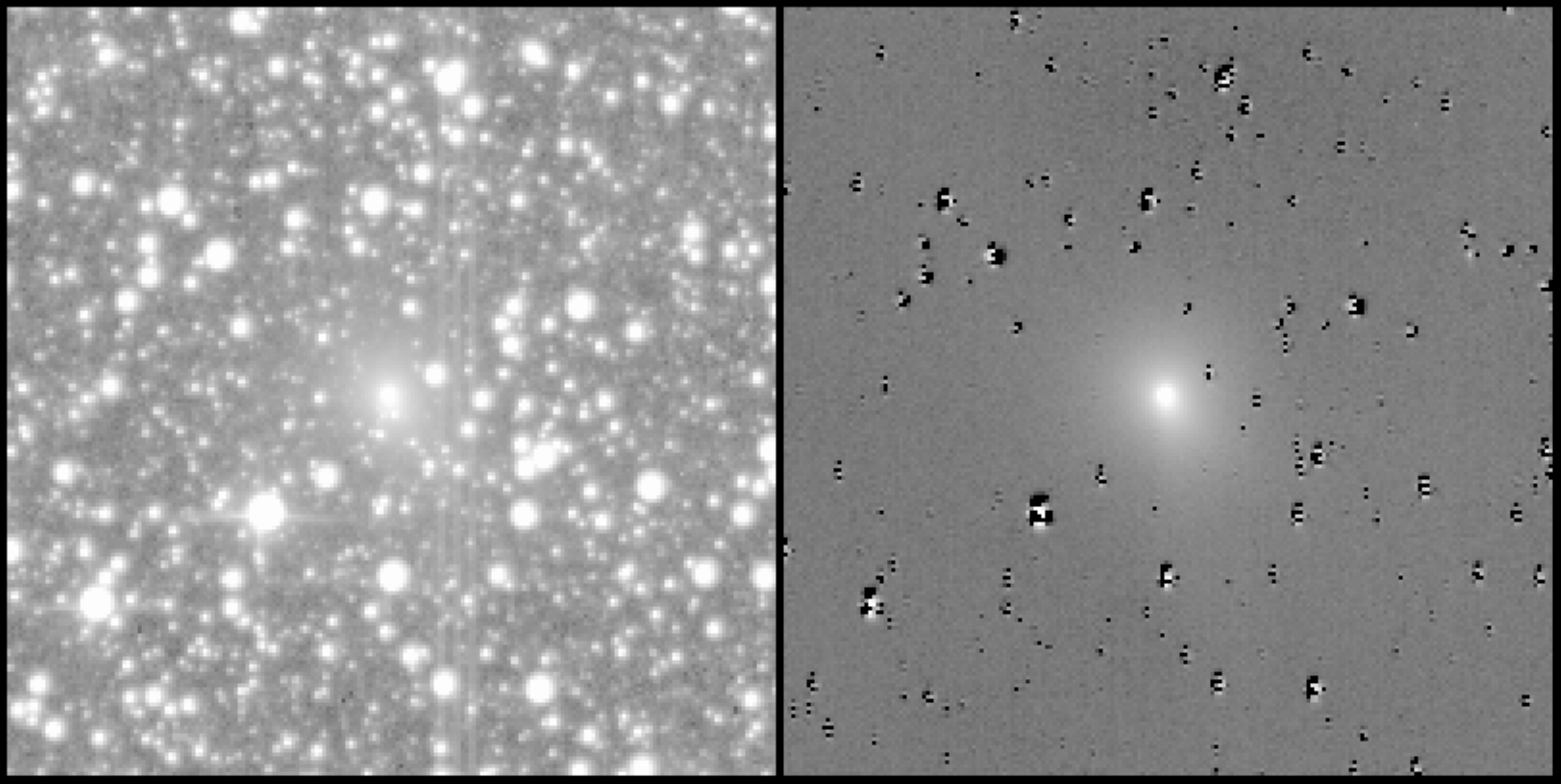}
\caption{Example of the star and background removal, showing the region of a
  sample frame (tess2018284152940-s0003-2-3-0123-s\_ffic\_sa) surrounding
  comet Wirtanen before (left) and after (right) the DIA cleaning
  process. The images \added{are 1.22$\times10^6$~km across and} are
    displayed with the same logarithmic scale.\label{fig:cleaned_image}}
\end{figure}

The FFI data are crowded with stars that must be removed to minimize their
interference with measurements of the moving comet and we adopted the TESS
user-provided {\em Difference Image Analysis} software (DIA;
\citealt{OelkersStassun:DIA}) to perform
this task. DIA is designed to highlight objects whose brightnesses vary with
time, and as an intermediate step in the reduction procedures the scattered
light (from the Earth, Moon, etc.) and background stars are removed from each
image.  We use these intermediate, cleaned images, an example of which is
shown in Figure~\ref{fig:cleaned_image}, for our comet analyses.  The
procedure does a good job removing the fainter stars and the wings of
brighter stars, but due to the undersampled PSF
\citep{VanderEtal:TESS_instrum}, uncorrelated residuals remain at the centers
of bright stars.  These residuals can still interfere with the comet
photometry, but at a significantly lower level than the stars in the original
images.

TESS pipeline products are calibrated to $\mathrm{e^\textrm{-}~s^{-1}}$,
which can be converted to TESS magnitudes (15,000
$\mathrm{e^\textrm{-}~s^{-1}}$ corresponds to $T=10.0$;
\citealt{VanderEtal:TESS_instrum}).  Using the Web TESS Viewing
Tool\protect{\footnote{\url{https://heasarc.gsfc.nasa.gov/cgi-bin/tess/webtess/wtv.py}}},
which requires an object's brightness in multiple bandpasses (we used
typical comet colors $V-J=1.47\pm0.17$ and $J-H=0.52\pm0.12$,
\citealt{HartmannEtal:cometcolor}; and $V-R=0.50\pm0.03$,
\citealt{Jewitt:colors}) we produced a rough conversion to $R$ magnitudes,.
The systematic uncertainty in $R$ is dominated by the uncertainties in the
comet colors, with a conservative estimate of $\pm0.3$~mag.  Relative
photometry from image to image, on the other hand, is more tightly
constrained, with errors $<0.017$~mag.  Flux units are computed using the
conversion coefficient for a $R=0$ star of
$2.18\times10^8~\mathrm{W~m^{-2}~\mu m^{-1}}$ \citep{BesselEtal:zeropoints}. 

\subsection{Lightcurve and Outburst} \label{sec:outburst}

\begin{figure}[hbt]
\plotone{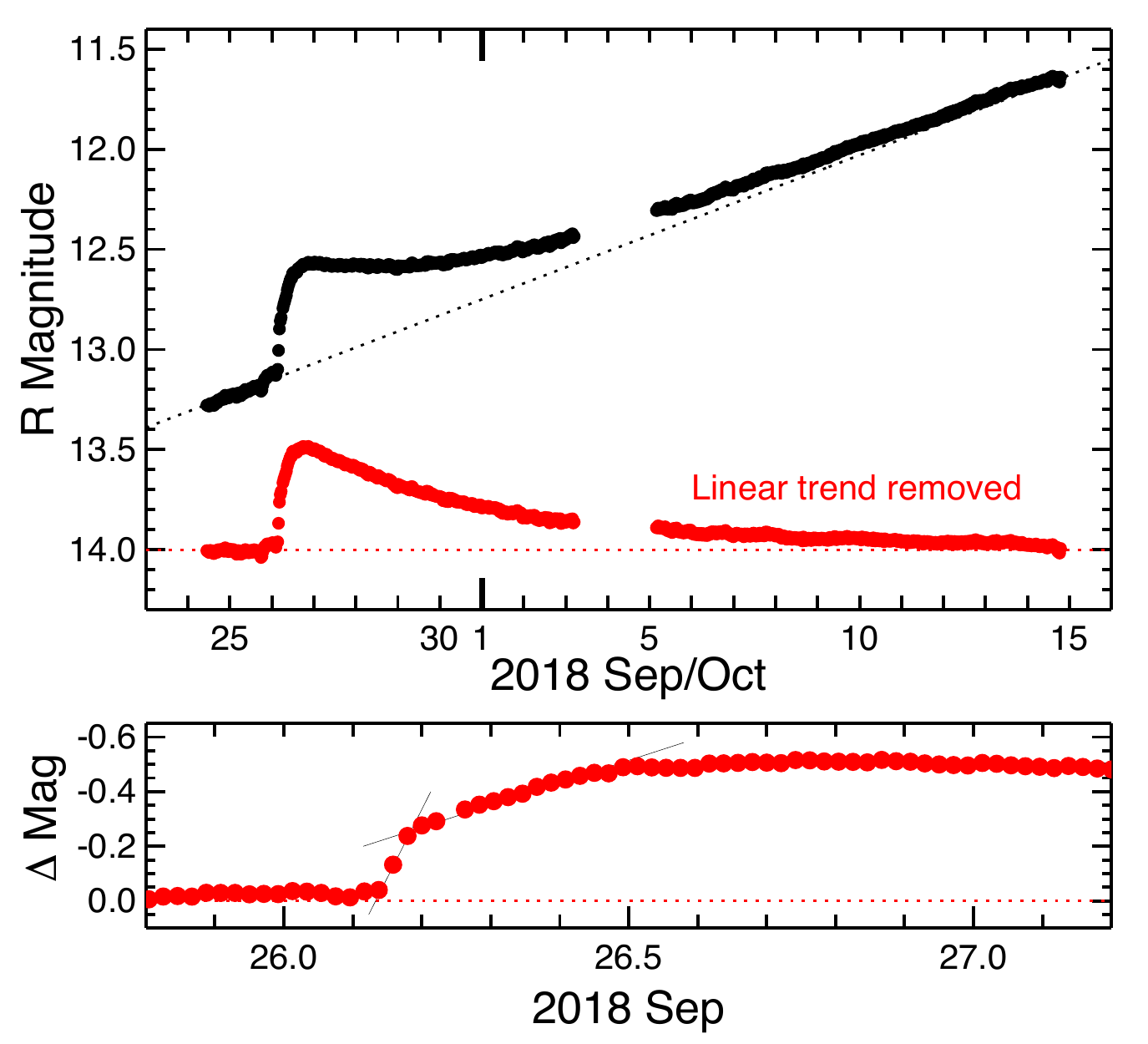}
\caption{Comet Wirtanen's lightcurve as measured in a 25,000~km
  \added{radius} aperture from the TESS images. The top panel shows the
  comet's secular brightening (black points) and reveals an outburst that
  begins at September~26.12. Removing an assumed linear baseline (dotted
  line) highlights the outburst behavior (red points).  The lower panel
  expands the section around the onset of the outburst, showing the rapid,
  hour-long jump (0.21~$\mathrm{mag~hr^{-1}}$), followed by a more gradual
  brightening (0.034~$\mathrm{mag~hr^{-1}}$) that peaks $\sim$8~hrs later.
  See Section~\ref{sec:reduction} for a discussion of the
  uncertainties. \label{fig:lightcurve}}
\end{figure}

Centering on the optocenter of the coma, we measured the brightness of the
comet in multiple apertures with fixed radii ranging from 15,000 to 40,000~km
at the distance of the comet.  To produce our lightcurve, we ultimately
selected the 25,000~km aperture, which is large enough to contain at least
3~PSFs at all comet distances, but minimizes the number of residual stars
crossing the aperture.  The resulting lightcurve, shown in
Figure~\ref{fig:lightcurve}, shows the comet continuously brightening over
the observation window, but more importantly, it reveals that TESS captured a
moderate-sized (0.5~mag) outburst---the only one to be reported during the
2018 apparition.  This event is likely to be the best \added{temporally}
characterized natural
outburst ever recorded, with the nominal pre-outburst behavior, the onset and
rise to the peak, and ultimately the falloff over the course of several
weeks, all being observed at a 30-minute cadence.  The observations show a
two-phase brightening, with a rapid $\sim$1~hr jump commencing on
September~26.12\added{$\pm$0.01}, followed by a more gradual increase that continued for
another $\sim$8~hr.  After peaking, the coma began a roughly exponential fading
that lasted 15-20~days.  Because the comet's pre-outburst brightening rate is
not well constrained, it is not clear whether or not the comet had returned
to its baseline level by the end of the observation window.

\begin{figure}[hbt]
\plotone{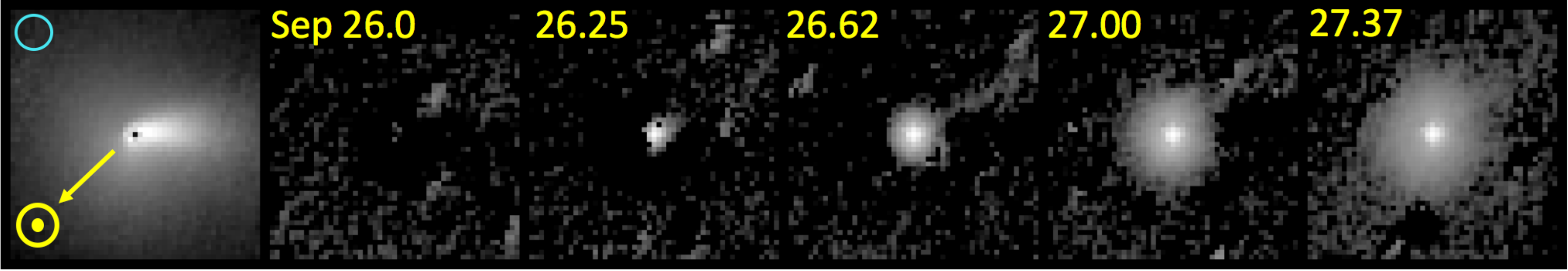}
\caption{Image sequence showing the outburst's effect on Wirtanen's coma.
  The first panel shows the pre-outburst morphology, enhanced by dividing out
  a 1/$\rho$ profile.  To show changes in the coma with time, panels 2--6 are
  enhanced by \replaced{dividing}{subtracting} out the (unenhanced) image
  from panel~1, \added{scaled by the linear trend shown in
    Figure~\ref{fig:lightcurve}}.  Panels~2 and~3 bracket the onset of the
  outburst (September~26.12) and~\replaced{3--5}{4--6} show the bright
  central condensation and the rapidly expanding gas cloud.  \deleted{The gap
    on the West side of the gas cloud is an artifact resulting from dividing
    out the bright dust tail.} Each panel is 400,000~km across, with North
  up and East to the left. \added{The light blue circle denotes a
    25,000~km radius aperture.}  \label{fig:outburst_image}}
\end{figure}

Following up on this discovery, we inspected the images for additional
information about the outburst.  We registered and coadded the data in 3~hour
blocks to improve the S/N, and then enhanced the post-outburst images by
dividing out a pre-outburst frame.  The resulting sequence (e.g., Figure
\ref{fig:outburst_image}) shows the rapid brightening of the central coma,
followed by an outflow of material that is roughly symmetric around the
nucleus. The leading edge of this outflow had a projected velocity
$\sim$800~$\mathrm{m~s^{-1}}$, suggesting that it was gas related to, and
possibly driving, the outburst.  Given the detector bandpass (600--1000~nm)
the gas was most likely dominated by CN, with emission bands at 914.1 and
787.3~nm \citep{schleicher:thesis}.  The gas contributes only a few percent
of the total brightness \replaced{increase, and shows no signature in
the lightcurve as it exits the photometric aperture}{within the aperture}.

\begin{figure}[hbt]
\plotone{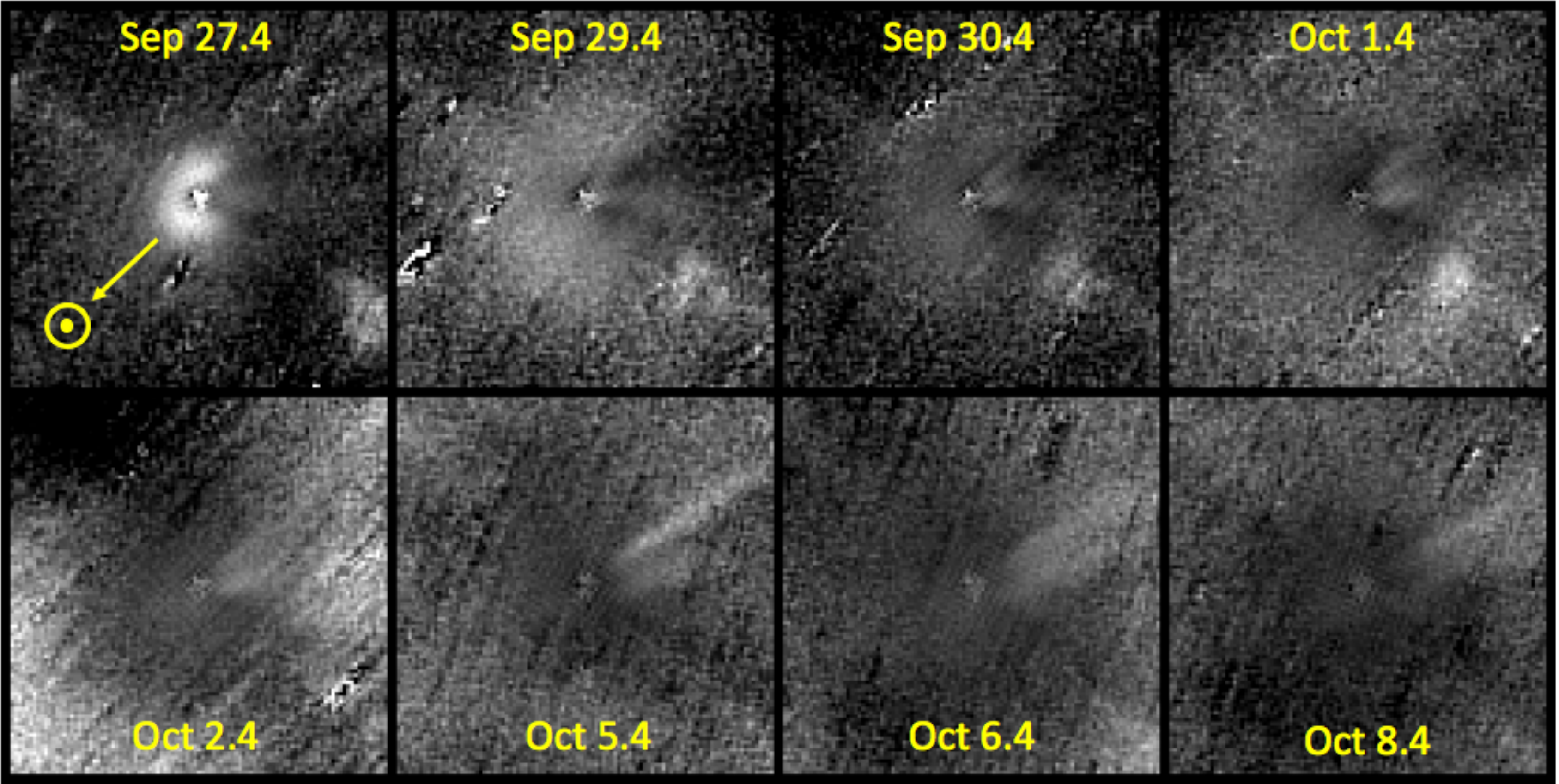}
\caption{\added{Image sequence showing the effects of Wirtanen's outburst on
    its dust tail.  The first panel highlights the outburst, followed by
    images that show a narrow tail forming in the anti-solar direction over
    subsequent days. Each panel consists of 8~hrs worth of images, centered
    on the stated date and rotated to align the sunward direction with that
    of the Sep~27 panel before coadding.  Images were enhanced by combining
    all images between September~28 and October~9 (again rotated to align the
    sunward direction) into a single temporally averaged template, which was
    then divided out of each of the eight coadded panels.  Images are 100
    pixels across, which ranges from 780,000~km on September~27 to 640,000~km
    on October~8.}  \label{fig:dusttail_image}}
\end{figure}

In images over the few days following the outburst, the rapidly moving gas
diffuses away, while, in contrast, the bright central peak (presumably dust)
remained highly concentrated, showing little radial expansion over the
following 15-20 days. This material left our photometric aperture primarily
when it dispersed down tail under the influence of radiation pressure
\added{(see Figure~\ref{fig:dusttail_image})}.  Thus, the optically dominant
grains ejected in this event must have been moving slowly, with expansion
velocities of a few tens of~m/s, (\added{consistent with those seen in
  outbursts in comet 49P/Arend-Rigaux and 67P/Churyumov-Gerasimenko}
\citep{EisnerEtal:arend,LinEtal:67Poutburst}) or else they would have left
the photometric aperture more rapidly than was observed.

\subsection{Dust Trail} \label{sec:trail}

\begin{figure}[hbt]
\plotone{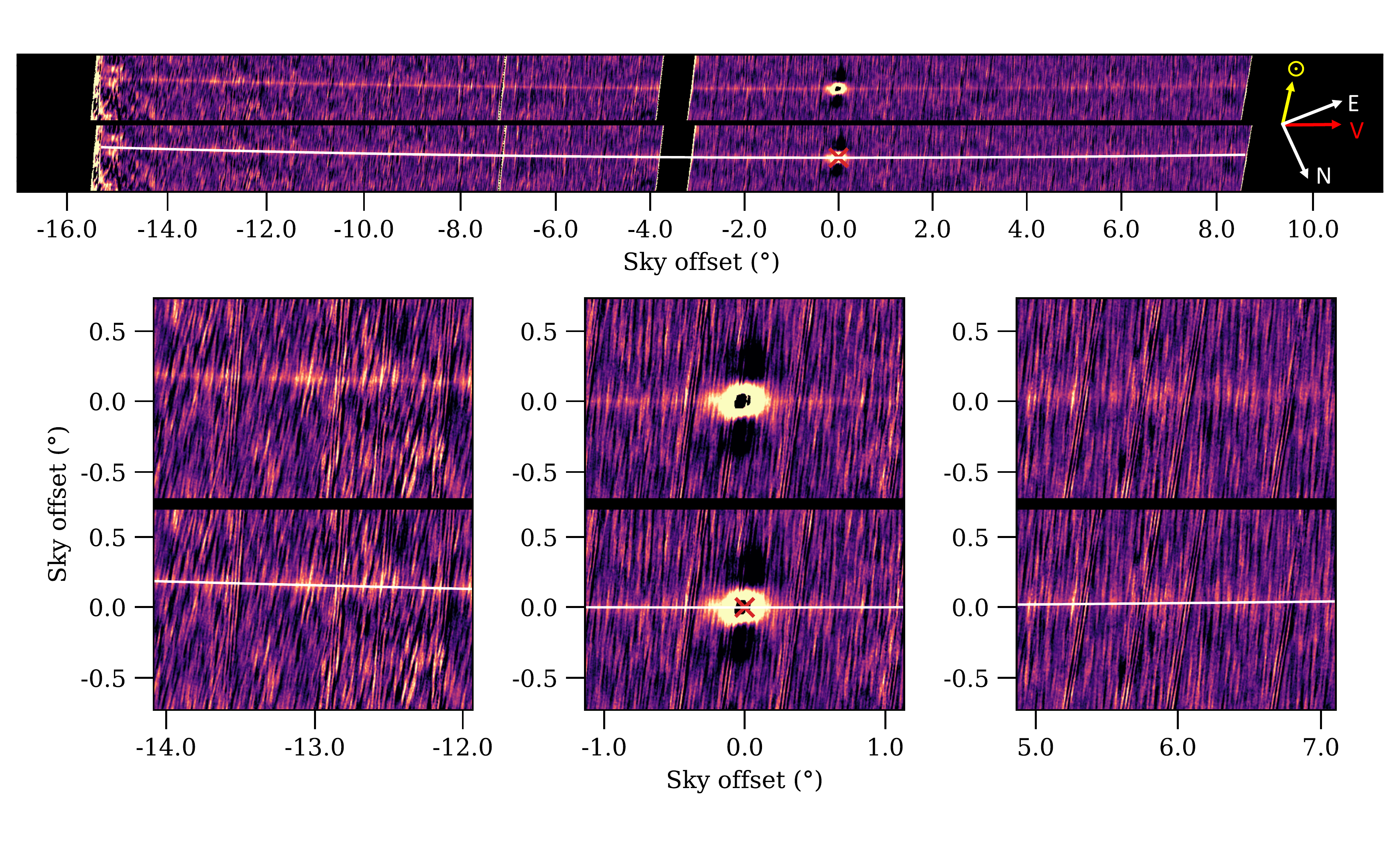}
\caption{Sample mosaic showing Wirtanen's dust trail, produced by combining
  240~images from the 5-day period from September~28 to October~3. The top
  panel shows the full mosaic, spanning CCDs~3 and~4 in Camera~2; sections at
  the nucleus and outer ends are expanded below.  In the lower frame of each
  image pair, a cross is located at the nucleus' position and the comet's
  orbit is overlain, showing that the trail is centered on the comet's path.
  The direction arrows in the inset are shown for the nucleus' position at
  the midtime of the five-day window.  \label{fig:trail}}
\end{figure}

Examination of the TESS images also revealed that Wirtanen has a previously
unreported dust trail.  Due to the presence of the comet itself, the DIA
software oversubtracts the background in the vicinity of the coma and trail.
To better investigate the trail, we grouped images into 5-day bins,
re-projected the raw frames into the comet's rest frame with the velocity
vector aligned, masked the bright stars, and then combined the images via the
median value at each pixel.  The resultant images were additionally filtered
to remove the CCD banding artifacts.  A sample mosaic is presented in
Figure~\ref{fig:trail}.

The dust trail extends in both directions beyond the~24\degr{} TESS
field-of-view, which covers a mean anomaly range ($\Delta M$)
of~$-1.6$\degr{} to~0.7\degr{}.  A Gaussian function fit to the width of the
trail at $\Delta M=-0.16$\degr{} (-1.7\degr{} on the sky) has a peak of
29.3~$\mathrm{mag~arcsec^{-2}}$ and a 1$\sigma$ width of
$(5.5\pm1)\times10^4$~km.  The width and optical depth
($\tau=7.0\times10^{-12}$) are consistent with the
faint end of the range of trails observed at infrared wavelengths
\citep{IshiguroEtal:DustTrails}, assuming a 4\% geometric albedo.

\section{Discussion} \label{sec:discussion}

Outbursts have been observed in comets for over a century
\citep[e.g.,][]{Hughes:outbursts,FiloChury:outburst}.  A number of proposals
have been put forth as trigger mechanisms
\citep{Hughes:outburstmech,GronkWeso:outburstrev}, though it is generally
believed that a single explanation is insufficient to account for all cases.
Similarly, little is known about the physical processes that govern the
behavior of the ejecta during the initial stages of the event.  The main
problem is that the majority of outbursts are detected only after they have
peaked in intensity and the quasi-exponential fading stage provides few
observational constraints on the conditions present during the explosive
phase.  The excellent coverage of the Wirtanen event allows us to use its
characteristics as a baseline model for the behavior of a moderate sized
outburst that can be compared and contrasted with other outbursts.  With this
information, we can begin exploring the physical processes at work.

\subsection{Characterization of Wirtanen's Outburst} \label{sec:character}

To begin this investigation, we can evaluate some potential trigger
mechanisms by looking at a comet's history.  Wirtanen is not known for
exhibiting outburst activity, and only three previous events were reported
since its 1948 discovery: 1991~October~7 (+16~days from perihelion, $<$1~mag,
\citealt{Kronk:cometog6}); 2002~September~25 (+29~days, $\sim$2~mag,
\citealt{Yoshida:website2002}, \citealt{Kidger:CGdust2}); and 2008~May~16
(+86~days, 2~mag, \citealt{Kidger:astroncirc2446}).  Including the 2018
outburst on September~26 (--81~days), there is no correlation in the orbital
position for these outbursts.  Furthermore, given the similar viewing
geometry on alternate apparitions, there are often observations from $\sim$11
years before or after the observed outbursts that rule out any events
occurring at those points on other orbits.  Thus, the lack of correlation
between outbursts and the comet's true anomaly or heliocentric distance,
combined with the general rarity of these events, suggests that a
persistently volatile region, as described by \cite{Miles:sw1_cryovol},
is not the cause of the Wirtanen events.

Next, we can characterize the scale of the event to provide context for
comparisons to other events.  The outburst produced a 0.5~mag brightness
increase.  For a more quantitative measurement, we convert this brightening
to an increase in reflective surface area, and then to a mass of the ejected
material.  We use the photometric analysis outlined by \cite{Jewitt:photom},
and assume that the brightness increase is dominated by sunlight reflected
off dust grains ejected into the coma. For an initial brightness $R=13.1$ and
a total brightening of 0.5~mag, we find an increase in the optical cross
section of ${\sim}2{\times}10^8$~m$^2$ (using the geometric parameters at the
start of the outburst, the Halley-Marcus composite dust phase function
compiled by
D.~Schleicher\protect\footnote{\url{https://asteroid.lowell.edu/comet/dustphase.html}},
and an assumed 4\% albedo).

Adopting a grain population with density 2000~$\mathrm{kg~m^{-3}}$, radii
ranging from 0.1~$\mu$m to 1~mm and a \added{differential size distribution 
with a} power-law index of $-3.5$, we find that the above cross section
converts to a total dust mass of ${\sim}3{\times}10^6$~kg.  Unfortunately,
the mass tends to be dominated by the largest particles and the dust
properties in the outburst are not well constrained, so changing the assumed
upper limit of the grain size and/or the power law index can alter the
computed mass by as much as two orders of magnitude for plausible
populations.  Thus, our best estimate is that the outburst likely produced
$10^5-10^8$~kg of material.  This is equivalent to a crater a few 10s of
meters in radius (for a bulk nucleus density of 500~$\mathrm{kg~m^{-3}}$),
which is consistent with the outburst-associated features seen on comet
9P/Tempel~1 \citep{BeltonEtal:cryovol}.  Pits known to be associated with
outbursts are also seen on comet 67P/Churyumov-Gerasimenko
\citep{VincentEtal:cometactivity}.  Although these features are smaller
($<$10~m) than that computed for Wirtanen, the related C-G outbursts were
also smaller, and were not detected from Earth.

An important factor in the Wirtanen outburst is that it is one of the first
outbursts whose rise in intensity is well documented, showing a two-phase
increase in brightening.  Given that the gas was measured to expand outward
at $\sim$800~$\mathrm{m~s^{-1}}$, while the dust exhibited significantly lower speeds, we
know that during the first eight hours following the initiation of the
outburst, even the leading edge of the gas cloud would not have escaped our
25,000~km aperture.  This means that the two-phase brightness profile must
reflect temporal characteristics of the outburst (e.g., the changes in slope
are not due to material leaving the aperture).  We suggest that the initial,
hour-long period of brightening represents the energetic phase in which gas
and dust were rapidly ejected into the coma.  The second, more gradual phase
of brightening is probably due to the continued expansion of the dust from an
initially dense state to an optically thin regime.  However, it is also
possible that this phase arises from other causes: a continued excavation, at
a much slower rate as the outburst subsides; temporary enhanced emission of
gas and dust produced by sublimation of newly exposed ices; or the increase
in reflective surface area as a small number of grains in the coma gradually
fragment into smaller particles.

\subsection{Comparison to Other Outbursts} \label{sec:compare}

Our literature search revealed only three other occasions in which
high-cadence coverage was obtained of the brightening phase of an outburst: a
2017 event in 29P/Schwassmann-Wachmann~1 (SW1), the 2007 outburst of comet
17P/Holmes, and the DI experiment in comet Tempel~1 in 2005.  As discussed
below, there are significant differences in the basic nature of each of these
events, but we can compare and contrast the observed behavior in the early
stages of each case, to explore whether similar mechanisms may be at work.

SW1 is known for frequent outbursts.  About half seem to be periodic
\citep{MilesEtal:sw1_anatomy}, suggesting that the nucleus has volatile-rich
hot spots that are triggered by diurnal heating cycles.  On 2017~July~2, SW1
experienced a $\sim$2~mag outburst and \cite{MilesEtal:sw1outburst} reported
on high-cadence photometry of this event.  Few results from these
measurements have been reported, but \replaced{for comparison, we used the
  same computation described above to estimate an}{if we adopt the same
  assumptions as used in the Wirtanen estimate above, we compute an} increase
in the SW1 optical cross section of ${\sim}10^{11}$~m$^2$.  \added{Because
  conditions are dramatically different in the two comets, this is likely a
  poor comparison, but it suggests that the SW1 outburst was} significantly
larger than \added{that} seen in Wirtanen.  Even so, the behavior \added{of the
  lightcurve} during the initial stages appears to be nearly identical in
character to the Wirtanen event, with a rapid rise for $\sim$0.5~hr, followed
by a more gradual increase that continued for at least another hour, when the
observations ended.  The fact that both events exhibit similar behavior, even
in their timescales, suggests that the \added{physcial} mechanisms
\replaced{at work}{governing the ejecta} are the same, even though the
magnitude of the events were dramatically different.

Comet Holmes experienced a 15~mag brightening---the largest outburst ever
recorded---around 2007~October~23.7 \citep{MontaltoEtal:holmes,
  SteveJewitt:holmes}.  The onset of the outburst was not captured, but the
42-hour duration of the rise allowed observers to capture the later stages of
the brightness increase \citep{TrigoEtal:holmes, HsiehEtal:holmes,
  LiEtal:holmes}.  This lightcurve shows that there was a change in the rate
of brightening around October~24.1, around half a day into the outburst,
where the rate became {\em steeper}, before inflecting and then peaking a day
later.  This steepening in slope was shown to be the result of a rapid
cascade of large dust grains fragmenting into successively smaller particles,
which dramatically increased the reflective surface area in the coma
\citep{HsiehEtal:holmes, SteveJewitt:holmes}.  In the days following the
outburst, the leading edge of the ejected dust cloud was measured to have a
velocity $\sim$550~$\mathrm{m~s^{-1}}$ \citep[e.g.,][]{TrigoEtal:holmes}, an
order of magnitude higher than the dust velocities in Wirtanen, which likely
reflects the enormous amount of energy released in the event.

Our final comparative example is the Deep Impact experiment at comet Tempel~1
(T1).  Although this was a man-made outburst, it was known to be the result
of an impact and should exhibit the same phenomena as a naturally-triggered
impact event.  Furthermore, because it was planned, it was intensely observed
from both DI and from the ground, allowing remote observations to be
connected to specific events seen by the spacecraft.  High cadence photometry
of the event shows a three-phase brightening \citep{MeechEtal:DI_campaign,
  FernandezEtal:Tempel1, MitchellEtal:tempel1, KuppersEtal:Rosetta_DI}.  The
T1 lightcurve starts with a very sharp increase for the first $\sim$1~min,
followed by 6~min of gradual brightening, and then another 10-15 minutes with
a somewhat steeper slope before it flattened out at $\sim$2~mag above its
starting brightness.  The first two phases mimic the behavior of Wirtanen and
SW1, while the later steepening is comparable to that seen in Holmes.  The
outburst started to fade $\sim$45 mins after impact.  The peak velocity of
the ejecta was ~$\sim$200~$\mathrm{m~s^{-1}}$.  It is interesting to note that details in the
observed phenomena, including changes in the brightness slope, peak
intensity, and timing of the observed features, vary somewhat depending on
the aperture size, wavelength, etc., and exploring these differences should
provide additional information for exploring the behavior of the ejecta.

As was suggested in Section~\ref{sec:character} the two-phase brightness
increase is related to the initial ejection of highly dense material that
then becomes optically thin as it expands outward.  The T1 data provide a
test of this conjecture.  \cite{KoloEtal:tempel1} used DI observations of the
ejecta in front of the comet's limb to measure the optical thickness of the
dust. They found that the cloud started out optically thick, but tended to
thin within a few seconds in most areas.  However, a few bands of material
remained optically thick for as long as a minute.  This time
\replaced{scale}{frame} corresponds to that of the steep segment of the
photometric profile, suggesting that the change in slope may indeed be due to
the expansion of the ejecta into an optically thin regime.  \added{The
difference in timescales between Wirtanen and T1 are likely to be due to
differences in velocity as well as the properties of the ejecta.}

The comparisons between the initial stages of these four outbursts reveal
both similarities and differences.  The similarities suggest that common
processes govern the physics, regardless of the cause or size of the event,
while the differences, such as the shorter timescale seen in T1 or the long
duration of the Holmes brightness increase, could represent characteristic
signatures that provide clues to the mechanisms involved.  Results for comet
Wirtanen also show the promise that TESS observations hold for cometary
science.

\acknowledgments
This paper includes data collected by the TESS mission, which are publicly
available from the Mikulski Archive for Space Telescopes (MAST). Funding for
the TESS mission is provided by NASA's Science Mission directorate.  

%

\vspace{5mm}
\facility{TESS}


\software{DIA \citep{OelkersEtal:DIA_CSTAR,OelkersStassun:DIA}, Astropy
  \citep{AstropyEtal:astropy, AstropyEtal:opensci}, SEP \citep{Barbary:sep}
}

\listofchanges

\end{document}